\documentclass[aps,prl,twocolumn,groupedaddress]{revtex4-1}

\newcommand{\be}{\begin{eqnarray}}
\newcommand{\ee}{\end{eqnarray}}
\newcommand{\ket}[1]{\ensuremath{\left| {#1} \right>}}
\newcommand{\bra}[1]{\ensuremath{\left< {#1} \right|}}

\newcommand{\Caplus}{\ensuremath{{^{40}}{\rm Ca}^{+} \,}}

\newcommand{\create}{\ensuremath{{\,\hat{a}^{\dagger}}}}
\newcommand{\destroy}{\ensuremath{{\,\hat{a}}}}

\newcommand{\splus}{\ensuremath{\hat{\sigma}_+}\,}
\newcommand{\sminus}{\ensuremath{\hat{\sigma}_-}\,}

\usepackage{latexsym}
\usepackage{graphics}
\usepackage{multirow}
\usepackage{textcomp}
\usepackage{siunitx}
\usepackage{todonotes}

\begin{document}
\title{Quantum harmonic oscillator state control in a squeezed Fock basis}

\author{D.~Kienzler}
\altaffiliation{Current Address: National Institute of Standards and Technology,
	Time and Frequency Division 688, 325 Broadway, Boulder, CO 80305, USA}
\author{H.-Y.~Lo}
\author{V.~Negnevitsky}
\author{C.~Fl{\"u}hmann}
\author{M.~Marinelli}
\author{J.~P.~Home }
\email[Corresponding author, Email: ]{jhome@phys.ethz.ch}

\address{Institute for Quantum Electronics, ETH Z\"urich, Otto-Stern-Weg 1, 8093 Z\"urich, Switzerland}


\begin{abstract}
We demonstrate control of a trapped-ion quantum harmonic oscillator in a squeezed Fock state basis, using  engineered Hamiltonians analogous to the Jaynes-Cummings and anti-Jaynes-Cummings forms. We demonstrate that for squeezed Fock states with low $n$ the engineered Hamiltonians reproduce the $\sqrt{n}$ scaling of the matrix elements which is typical of Jaynes-Cummings physics, and also examine deviations due to the finite wavelength of our control fields. Starting from a squeezed vacuum state, we apply sequences of alternating transfer pulses which allow us to climb the squeezed Fock state ladder, creating states up to excitations of $n = 6$ with up to 8.7 dB of squeezing, as well as demonstrating superpositions of these states.  These techniques offer access to new sets of states of the harmonic oscillator which may be applicable for precision metrology or quantum information science.
\end{abstract}

\pacs{pacs}
\maketitle

The control of quantum harmonic oscillators has played a prominent role in the development of quantum state control \cite{13Haroche, 13Wineland}. A single quantum oscillator provides access to a Hilbert space with a dimension which increases rapidly as the oscillator energy increases. It is also an example of a system which has a natural
transition from the quantum to the classical regimes. One of the primary methods for performing control and measurement of quantum harmonic oscillator states is by coupling the oscillator to a single spin using a Jaynes-Cummings Hamiltonian
\be
\hat{H}_{\rm JC} = \hbar\Omega/2 (\hat{a}^{\dagger} \sminus e^{i \phi} + {\rm h.c.})
\ee
where $\Omega$ and $\phi$ are real constants, $\create$ and $\destroy$ are the creation and annihiliation operators of energy quanta for the oscillator, and $\sminus = \ket{\downarrow}\bra{\uparrow}$ with $\splus = \hat{\sigma}_-^{\dagger}$. $\ket{\uparrow}$, $\ket{\downarrow}$ are energy eigenstates of the spin. The Jaynes-Cummings Hamiltonian arises naturally for Cavity-QED systems \cite{BkHaroche} and can be implemented straightforwardly with trapped ions by using a laser to resonantly drive a motional sideband of an internal state transition \cite{13Wineland, 96Meekhof, 03SchmidtKaler}. For an ion starting in one of the basis elements  $\ket{\downarrow}\ket{n}$ where $\ket{n}$ are the energy eigenstates of the oscillator, evolution as a function of the duration $t$ of the Hamiltonian $\hat{H}_{\rm JC}$ results in Rabi oscillations between the two states $\ket{\downarrow} \ket{n} \leftrightarrow \ket{\uparrow}\ket{n -1}$ which can be viewed as a rotation $R(\theta, \phi) = \cos(\theta/2) \hat{I} + i \sin(\theta/2) (\cos(\phi)\hat{s}_x - \sin(\phi)\hat{s}_y)$ where $\theta = \Omega \sqrt{n} t$ and $\hat{I}, \hat{s}_x, \hat{s}_y$ are Pauli operators which act in the basis $\{\ket{\downarrow}\ket{n}, \ket{\uparrow}\ket{n-1}\}$.  These Rabi oscillations can be observed by making projective measurements on the spin states of the ion as a function of the duration of the applied Hamiltonian. An important feature is that the Rabi frequency of the oscillations scales with the matrix element $ \bra{n - 1} \destroy \ket{n} = \sqrt{n}$. This has played an important role in the diagnosis of energy distributions of various well-known oscillator states which are not energy eigenstates \cite{13Haroche}.  The use of coupling to a two-state system isolates pairs of states of the oscillator, which simplifies the dynamical evolution and allows simple prescriptions for creating arbitrary superpositions of states \cite{96Law, 03BenKish, 09Hofheinz}.

Although the Jaynes-Cummings Hamiltonian arises naturally in the light-matter interaction, similar physics can be observed for any Hamiltonian of the form
\be
\hat{H}_- = (\hbar\Omega_-/2) [\hat{K} \splus e^{i\phi}+ {\rm h.c.}  ] \ ,
\ee
where $\hat{K}$ is a non-Hermitian operator acting on the oscillator for which the commutation relation $[\hat{K}, \hat{K}^\dagger ] = 1$ holds. These operators then create and annihilate excitations on a Fock state ladder which is not the energy eigenbasis. Creation and annihilation operators of this type can be obtained by making a unitary transformation such that $\hat{K} \equiv \hat{U}\hat{a}\hat{U}^\dagger$, with the corresponding Fock state ladder given by $\hat{U}\ket{n}$. Since $\hat{U}$ is unitary, the commutation relation is preserved and the matrix elements follow the same scaling with excitation number $n$ as for the energy eigenstates, namely that
\be
\bra{n-1} \hat{U}^\dagger \hat{K} \hat{U} \ket{n}  = \sqrt{n} \ . \label{eq:sqrtn}
\ee
These observations allow all elements of control devised for the Jaynes-Cummings Hamiltonian to be directly applied in the new basis, so long as the operators $\hat{K}$ can be implemented in the laboratory. Similarly, $\hat{H}_+ =  (\hbar \Omega_+/2) [\hat{K}^\dagger \splus  e^{i\phi} + {\rm h.c.} ]$ is analogous to the anti-Jaynes-Cummings Hamiltonian. One class of unitary transformations which can be easily generated in the laboratory using trapped ions involve combinations of quadrature squeezing and displacements. These lead to operators of the form $\hat{K} = \mu \destroy + \nu \create - \alpha$, and thus both $\hat{H}_-$ and $\hat{H}_+$ can be produced by using multi-chromatic laser fields which simultaneously drive both the carrier transition $\ket{\uparrow} \leftrightarrow \ket{\downarrow}$ and the first motional sidebands $\ket{\uparrow}\ket{n} \leftrightarrow \ket{\downarrow}\ket{n \pm 1}$. In earlier work, we used this observation to demonstrate reservoir engineering for the creation of squeezed and displaced vacuum states, which can be viewed as cooling to the ground state of a given engineered basis. We also showed how the use of $\hat{H}_+$ provides a simple diagnosis that the ground state had been produced \cite{15Kienzler}. More recently, we used the predicted $\sqrt{n}$ scaling of Rabi oscillation frequencies produced by $\hat{H}_+$ to reconstruct Schr\"odinger's cat states of a motional oscillator which were beyond the range of standard methods which use only an energy-eigenstate decomposition \cite{16Kienzler}.

In this Letter, we demonstrate coherent control of a trapped-ion oscillator on an engineered squeezed Fock state ladder \cite{56Plebanski,85Satyanarayana,90Kral,97Nieto}. We apply sequences of pulses on both $\hat{H}_-$ and $\hat{H}_+$ to climb the state ladder, starting from the relevant ground state. We verify that the generated states are squeezed Fock states by extracting their number state distribution in the energy eigenbasis, and demonstrate that the Jaynes-Cummings analogy is valid by verifying the $\sqrt{n}$ scaling of the Rabi frequency with respect to the excitation of the basis states. We observe deviations from $\sqrt{n}$ scaling which are consistent with the finite Lamb-Dicke parameter used in our experiments. Finally, following the methods of Law and Eberly \cite{96Law} and earlier experimental work \cite{03BenKish} performed in the energy eigenstate basis, we use the engineered Hamiltonians to produce a coherent superposition of two Fock states of the squeezed state basis, and verify the key features of this state.

The experiments work with the axial mechanical oscillations of a single \Caplus ion trapped in a micro-fabricated ion trap with a secular frequency of $\omega_z/(2 \pi)$ =  \SI{2.07}{\mega\hertz}.  Spin-motion Hamiltonians make use of a two-level pseudospin defined as $\ket{\downarrow} \equiv \ket{L = 0, J = 1/2, M_J = 1/2}$ and $\ket{\uparrow}\equiv\ket{L'=2, J' = 5/2, M'_J = 3/2}$ . Coupling between spin and motion is implemented using a narrow-linewidth laser at  \SI{729}{\nano\meter} with a $k$-vector at 45 degrees to the axial oscillation direction, resulting in a Lamb-Dicke parameter of $\eta \approx 0.05$.
Working in a squeezed Fock basis requires implementing Hamiltonians generated from $\hat{H}_{\rm JC}$ using the squeezing operator $\hat{U} = \hat{S}(\zeta) \equiv \exp{[(\zeta \create^2 - \zeta^* \destroy^2)/2]}$ where $\zeta = r e^{i\phi}$ with $r$ and $\phi$ real numbers which relate to the magnitude and phase of squeezing. To implement a Hamiltonian of the Jaynes-Cummings type, we apply $\hat{H}_-$ with $\hat{K} = \hat{S}(\zeta)\destroy\hat{S}^\dagger(\zeta) = \cosh(r) (\destroy + \tanh(r)  e^{i\phi_s}\create)$. This can be produced by simultaneously driving the blue motional sideband ($\hat{H}_b \equiv (\hbar \Omega_b/2)[\create \hat{\sigma}_+  e^{i\phi_b}+ {\rm h.c.}]$) and red motional sideband ($\hat{H}_r \equiv (\hbar \Omega_r/2)[\destroy \hat{\sigma}_+ e^{i\phi_r} + {\rm h.c.}]$), and choosing the Rabi frequency ratio $\Omega_b/\Omega_r = \tanh(r)$ and a well-defined relative phase $\phi_s = \phi_b-\phi_r$. In what follows, the relevant state ladder will be written as $\ket{\zeta, n} \equiv \hat{S}(\zeta)\ket{n}$.

The initial step of each experiment involves initialization of the oscillator into the ground state of the squeezed basis by reservoir engineering \cite{15Kienzler, 96Poyatos}, which can be viewed as a modification of sideband cooling implemented by a combination of applying the engineered Hamiltonian $\hat{H}_-$ and optical pumping from $\ket{\uparrow}$ to $\ket{\downarrow}$. In the ideal case, the oscillator steady state of this process is the squeezed vacuum state $\ket{\zeta, 0}$. We verify the amount of squeezing by measuring the energy eigenstate occupations, and fitting these with the expected form for a squeezed state. This measurement is performed by switching on the blue-sideband Hamiltonian for a probe time $t_p$, and subsequently measuring the spin state. The probability of observing spin $\ket{\downarrow}$ in this measurement is given by
\be
P(\downarrow, t_p) = \frac{1}{2}\sum_k p(\ket{k}) \left( 1 + \gamma(t) \cos(\Omega_{k, k + 1} t_p) \right) \ , \label{eq:spinpops}
\ee
where  $p(\ket{k})$ is the probability that the motion started in the $k$th energy eigenstate state prior to the probe pulse. In the Lamb-Dicke regime the motion-dependent Rabi frequencies are given by $\Omega_{k, k+1} = \sqrt{k}\Omega_{b}$. For the data presented below we used $r \simeq 1$, which for an ideal squeezed vacuum state would correspond to a reduction in the quadrature variance in the squeezed direction of 8.7~dB.

In order to prepare the first excited state of the squeezed Fock state ladder $\ket{\zeta, 1}$, we use $\hat{H}_+$ to drive Rabi oscillations on the transition $\ket{\downarrow}\ket{\zeta, 0}\leftrightarrow \ket{\uparrow}\ket{\zeta, 1}$. At time $t_1 \simeq \pi/\Omega_+$, the spin state is fully inverted, and simultaneously the motional state is transferred from $\ket{\zeta, 0}\rightarrow \ket{\zeta, 1}$ (finite switching times of optical components mean that the inversion time $t_1$ is not exactly equal to $\pi/\Omega_+$). With the ion in the state $\ket{\uparrow}\ket{\zeta, 1}$, we then proceed to create $\ket{\zeta, 2}$ in an analogous fashion. Since the ion starts in the $\ket{\uparrow}$ state, we now apply $\hat{H}_-$ in order to drive the transition $\ket{\uparrow}\ket{\zeta, 1} \leftrightarrow \ket{\downarrow}\ket{\zeta, 2}$. By calibrating the time for which a full spin transfer is achieved and fixing the pulse length to this time, we create the squeezed Fock state $\ket{\zeta, 2}$.  By alternating between transfer pulses performed on $\hat{H}_+$ and $\hat{H}_-$, we are able to climb the Jaynes-Cummings state ladder as was performed previously for energy eigenstates \cite{96Meekhof}.

\begin{figure}[ht!]
\resizebox{0.49 \textwidth}{!}{\includegraphics{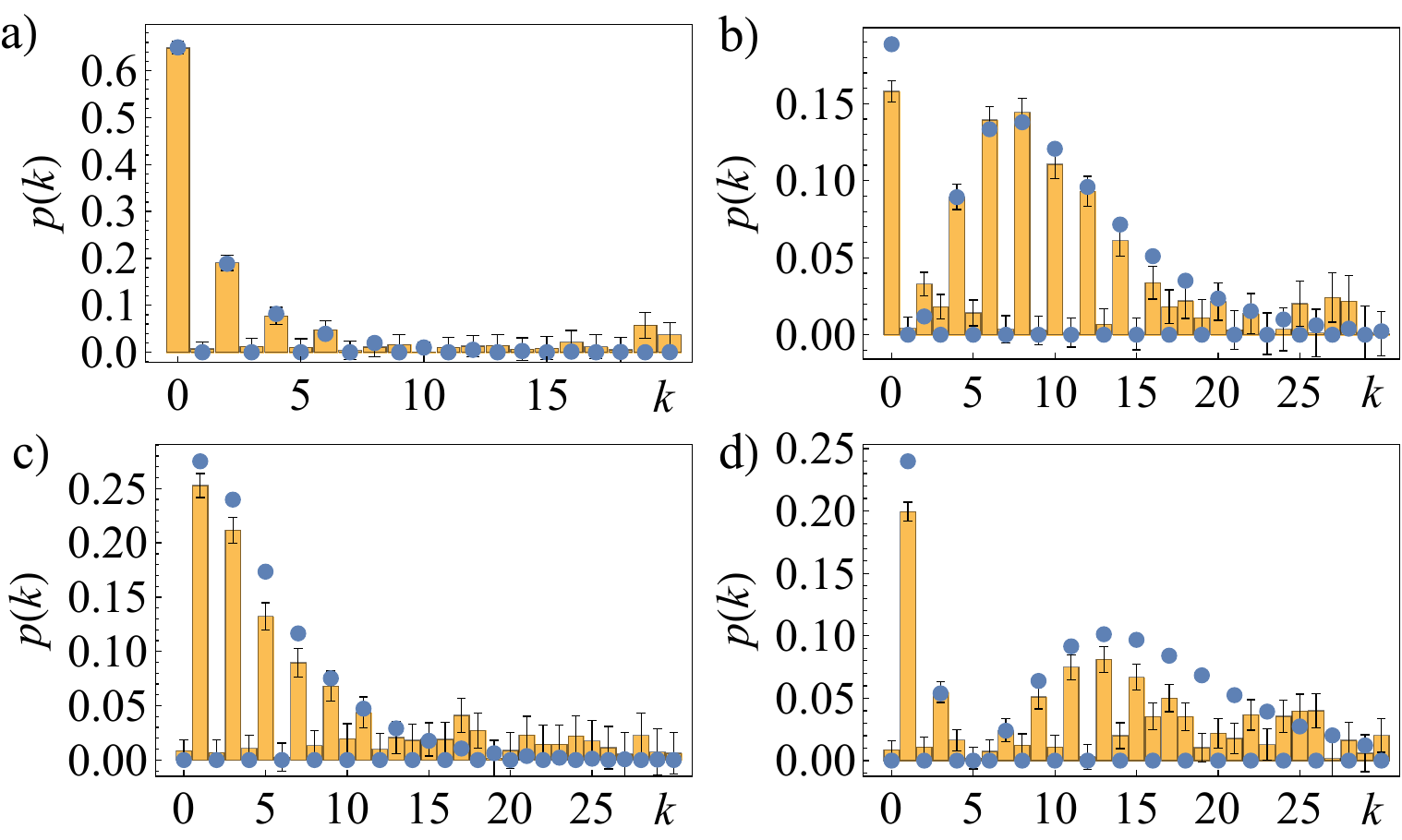}}
\caption{Energy eigenstate populations of a) \ket{\zeta, 0}, b) \ket{\zeta, 1}, c) \ket{\zeta, 2}, d) \ket{\zeta, 3}. The heights of the yellow bars indicate the probability of finding the ion in the $n$th energy eigenstate, with error bars included as standard error on the mean extracted from fits. The blue points indicate the expected populations for the squeezed Fock states with $r = 1$. As can be seen from the increased error bars, for Fock states above $k = 16$ it is difficult to separate contributions of neighbouring states. This is due to the limited number of Rabi oscillations which we can perform in our experiment and the reduced separation between the Rabi oscillations on neighboring energy eigenstates. }
\label{fig:projection}
\end{figure}

To verify that the states we produce are consistent with squeezed Fock states, we extract the energy eigenstate populations for each using blue-sideband Rabi oscillations. The extracted motional state populations for states up to $k = 30$ and $n$ up to 3 with $r = 1.00\pm0.03$ are shown in figure \ref{fig:projection}, showing agreement with the predictions for the squeezed Fock states $\ket{\zeta, n}$ for $r = 1$. A clear feature of the squeezed Fock states are their parity $\langle \hat{P}\rangle  = \sum_k (-1)^k p(\ket{k})$, for which we obtain values $0.99\pm0.14$, $-1.02\pm0.09$, $0.87\pm0.07$ and $-0.45\pm0.06$ for $n = 0 ,1,  2, 3$ respectively. These are negative for the odd squeezed number states and vice versa. This is expected since the squeezing operator preserves parity. For higher $n$, the agreement between the extracted motional state populations and the theoretical expectations becomes less good. We think that this is primarily a problem of the reconstruction method. The blue-sideband technique relies on separating out different oscillation frequencies in the time evolution of the spin population versus time. For higher $k$, the approximate $\sqrt{k}$ scaling of the matrix elements mean that neighboring Fock states produce similar Rabi frequencies, which require a large number of Rabi oscillations to be resolved. This is challenging because of fluctuations in the laser intensity driving the transition, which limit the number of oscillations for which a high oscillation contrast can be observed.

\begin{figure}[ht!]
\resizebox{0.49 \textwidth}{!}{\includegraphics{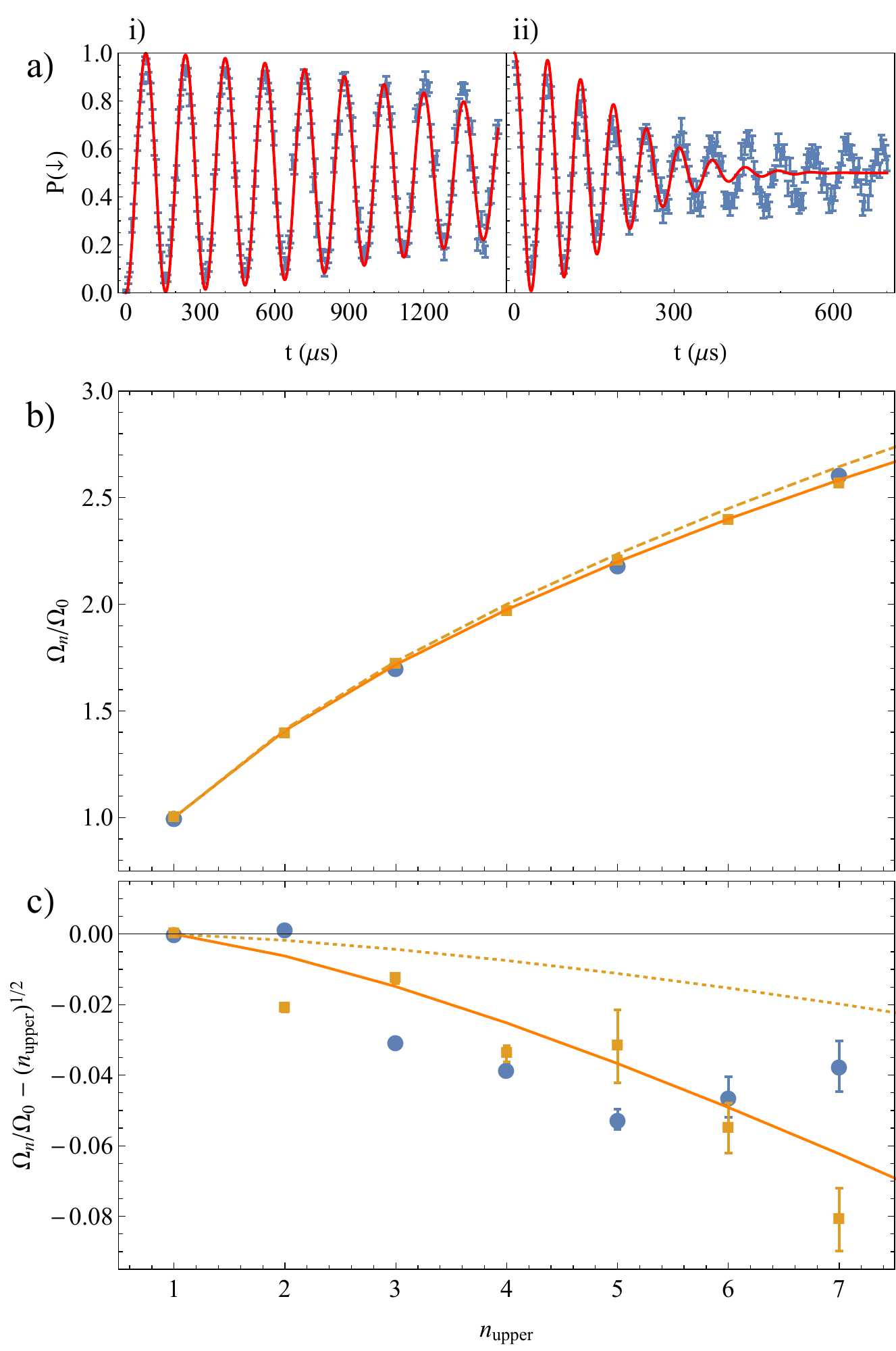}}
\caption{Rabi frequency scaling in the squeezed Fock basis. a) Example data with Rabi oscillation fits using a Gaussian decay to the spin populations vs. time for  i) the $\ket{\uparrow}\ket{\zeta, 0}\leftrightarrow\ket{\downarrow}\ket{\zeta, 1}$ and ii) the $\ket{\downarrow}\ket{\zeta, 5}\leftrightarrow\ket{\uparrow}\ket{\zeta, 6}$ transitions. b)  The measured ratio $\Omega_{n, n-1}/\Omega_0$  obtained from similar data for squeezed Fock states up to $n = 7$ with $r = 1$. Squares are measurements made using $\hat{H}_{+}$ and circles for $\hat{H}_-$. Theoretical curves are also plotted for $\sqrt{n}$ (dashed line) and for the modified theory taking into account the finite Lamb-Dicke parameter (solid line). c) A plot of measured $\Omega_{n, n-1}/\Omega_0 - \sqrt{n}$, showing the deviations from the $\sqrt{n}$ behaviour. The statistical error bars do not account for the deviations of the measurement from the theory (solid line), indicating that the error sources are primarily systematic. In simulations we see that a finite detuning between half the difference frequency of the lasers and the trap frequency increases the Rabi frequency (see supplemental material for more details). For comparison, the dotted curve indicates the matrix elements for the $n$th state of the energy eigenbasis for the same Lamb-Dicke parameter.}
\label{fig:RabiFreq}
\end{figure}

An important feature of Jaynes-Cummings physics is the $\sqrt{n}$ scaling of the matrix elements given in equation \ref{eq:sqrtn}, which is proportional to the Rabi oscillation frequency of transitions between the neighboring squeezed Fock states. In order to measure this in our experiments, we use a pulse of length $t \gg \pi/\Omega_{\pm}$ to drive many cycles of Rabi oscillation for each of the transitions $\ket{\downarrow(\uparrow)} \ket{\zeta, n}\leftrightarrow \ket{\uparrow(\downarrow)} \ket{\zeta, n +1}$, and extract the Rabi frequency by fitting the function $(1 + e^{-\gamma^2 t^2}\cos(\Omega t))/2$ with $\gamma$ and $\Omega$ floated parameters. For each transition between neighboring oscillator states, Rabi oscillations are possible either using $\hat{H}_+$ or $\hat{H}_-$. Since the laser settings used for both Hamiltonians are different, it is challenging to set exactly the same Rabi frequency for $\Omega_-$ and $\Omega_+$ respectively. To overcome this problem, we extract the Rabi frequency for both $\hat{H}_+$ and $\hat{H}_-$ for each motional state transition. For $\ket{\downarrow} \ket{\zeta, 0}\leftrightarrow \ket{\uparrow} \ket{\zeta, 1}$ the measurement of $\hat{H}_+$ is performed as before. Measurement of $\hat{H}_-$ on $\ket{\uparrow} \ket{\zeta, 0}\leftrightarrow \ket{\downarrow} \ket{\zeta, 1}$ is performed by driving Rabi oscillations after first transferring population from $\ket{\downarrow}\ket{\zeta, 0}$ to $\ket{\uparrow}\ket{\zeta, 0}$ using a carrier $\pi$-pulse. For measurements of the Rabi frequencies between higher motional levels, sequences of alternated transfer pulses on $\hat{H}_+$ and $\hat{H}_-$ are used to prepare the starting state prior to driving the measured Rabi oscillations. Two examples of observed Rabi oscillations are plotted alongside the scalings extracted from data in figure \ref{fig:RabiFreq}. Also plotted in figure \ref{fig:RabiFreq} b) is a theoretical curve for $\sqrt{n}\Omega_s$ with $\Omega_s$ obtained from the mean of the $\ket{\downarrow(\uparrow)} \ket{\zeta, 0}\leftrightarrow \ket{\uparrow(\downarrow)}\ket{\zeta, 1}$ Rabi frequencies for each Hamiltonian. While the trend of the Rabi oscillations is similar to the expected $\sqrt{n}$ scaling, the results do not agree to within experimental errors. The deviation of the data from the $\sqrt{n}$ theory is shown in figure \ref{fig:RabiFreq} c), along with a curve which accounts for the resonant term in the ion-light interaction to all orders of the Lamb-Dicke parameter. The latter clearly shows better agreement with the experimental data. For comparison, the standard scaling of the Lamb-Dicke correction for the energy eigenstates is also plotted in figure \ref{fig:RabiFreq}, showing that due to the higher energies of the squeezed Fock states for a given $n$, the correction factor is larger.

One notable feature that is observable as we ascend the ladder of states is that the coherence of the Rabi oscillations is reduced. This is an effect which might be expected both due to heating \cite{00Turchette2, 15Kienzler} or due to trap frequency fluctuations. At $n>3$ we observe that the data shows a persistent low amplitude oscillation which extends beyond the decay envelopes  which fit the early time part of the data for both Gaussian and exponential models of decay. Such an effect can be seen clearly in the data for the $n = 5 \leftrightarrow 6$ flopping given in figure \ref{fig:RabiFreq} a). We see that this behaviour is consistent with models which include a frequency offset of the difference frequency of the red and blue-sideband laser beams from the trap frequency, plus fast trap frequency fluctuations of a few Hertz.  As is discussed in more depth in the supplemental material, the sensitivity of the squeezed Fock states depends both on the level of squeezing and the excitation number, and is significantly enhanced for higher $n$.

\begin{figure}[ht!]
\centering
\resizebox{0.49 \textwidth}{!}{\includegraphics{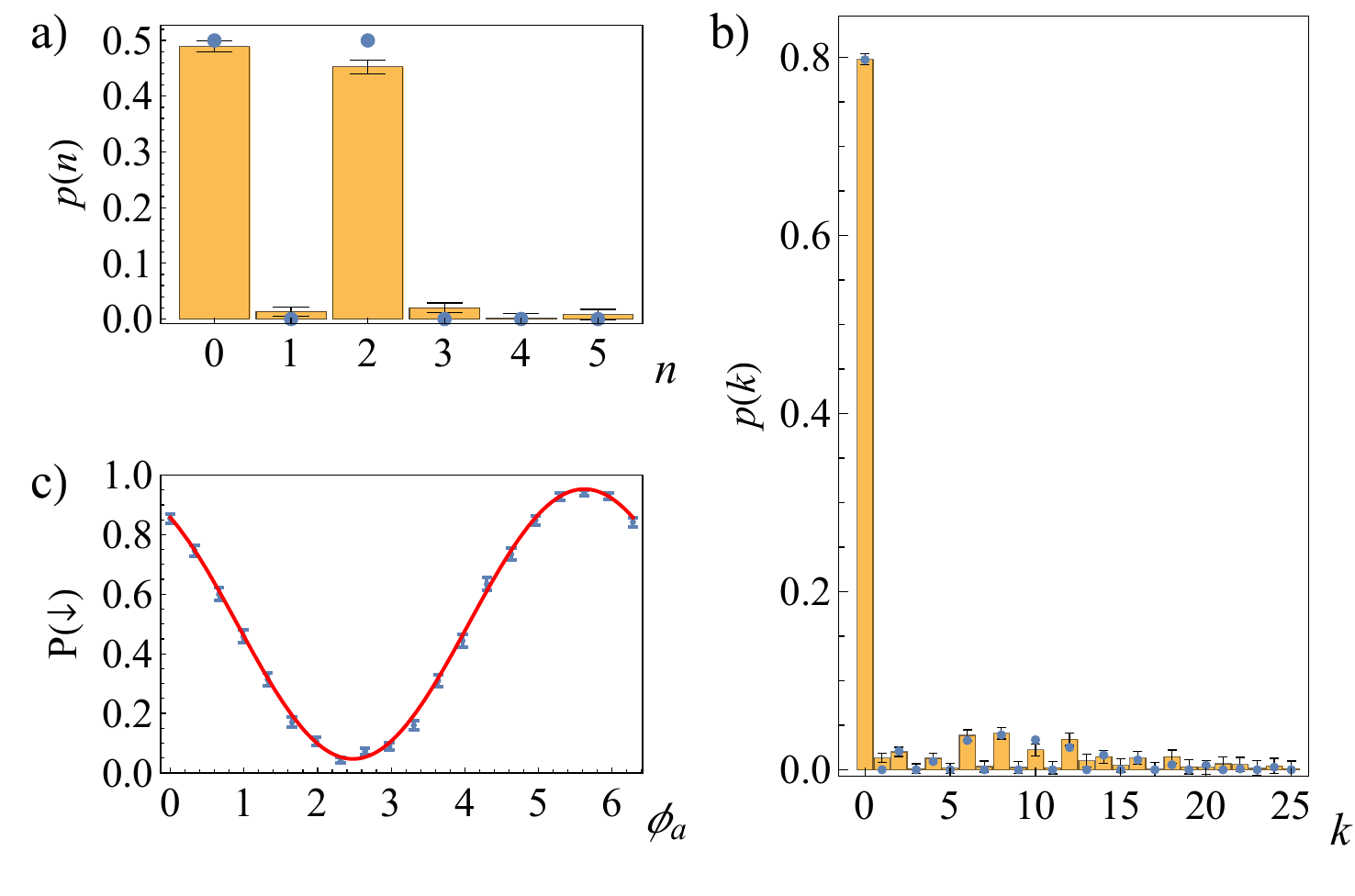}}
\caption{Measurements performed on a superposition of squeezed-Fock states. a) Population analysis in the squeezed-Fock basis, with $p(n)$ close to 0.5 for both $n = 0$ and $n = 2$. The bar heights and error bars are extracted from fits to Rabi oscillation data with the $\hat{H}_+$ Hamiltonian. Blue points are the expected values for the desired superposition.  b) Energy eigenstate population analysis, showing that the primary occupied levels are those with even parity. c) Population of the $\downarrow$ internal state as a function of the phase $\phi_a$ of a red-sideband analysis pulse, indicating the phase coherence between the superposed states. The contrast gives a lower bound on the fidelity of the superposition state. }
\label{fig:superposition}
\end{figure}

The toolbox of Jaynes-Cummings physics demonstrated above allows not only eigenstates of the chosen basis to be created, but also superpositions of those eigenstates. In order to demonstrate this, we create an equal superposition of $\ket{\zeta, 0}$ and $\ket{\zeta, 2}$. Starting from the squeezed ground state, we apply a pulse of duration $t_1/2$ on $\hat{H}_+$, resulting in the rotation $R_{+}(\pi/2, 0)$ acting now on the squeezed Fock states which creates the equal superposition $\left(\ket{\downarrow}\ket{\zeta, 0} + \ket{\uparrow}\ket{\zeta, 1}\right)/\sqrt{2}$. This is followed by a pulse with duration $t_2$ using $\hat{H}_-$, which performs a $\theta = \pi$ rotation which transfers $\ket{\uparrow}\ket{\zeta, 1}\rightarrow\ket{\downarrow}\ket{\zeta, 2}$, thus producing the desired superposition state $\ket{\psi_s} = \ket{\downarrow}(\ket{\zeta, 0} + e^{i \phi_s} \ket{\zeta, 2})/\sqrt{2}$. The phase $\phi_s$ is imposed by the laser phase, along with systematic AC Stark shifts during the creation of the state. We analyze $\ket{\psi_s}$ using three methods. First, we obtain the motional state occupations in the squeezed Fock state basis, by repumping the spin and subsequently measuring the population of the $\ket{\uparrow}$ state as a function of the duration $t_p$ of a pulse on $\hat{H}_+$. Experimental results are shown in figure \ref{fig:superposition} a), showing clearly that the dominant motional eigenstates are $\ket{\zeta, 0}$ and $\ket{\zeta, 2}$. In figure \ref{fig:superposition} b) we show the results of the number state decomposition in the energy eigenstate basis using the blue sideband, which is sensitive to the relationship between the squeezing axis and the phase relationship of the two states. Finally we examine the phase coherence between the two states of the superposition, by applying sequentially a $\theta = \pi$ rotation on $\hat{H}_-$ followed by a $\theta = \pi/2$ pulse on $\hat{H}_+$ with a phase $\phi_a$. We measure the final spin state as a function of the phase $\phi_a$. In an ideal realization the probability to find the ion in $\ket{\downarrow}$ would follow $P(\downarrow) = (1 + \cos(\phi_a - \phi_s))/2$. Data from such a scan are shown in figure \ref{fig:superposition} c), and show a sinusoidal oscillation as a function of phase, but with a reduced contrast which we extract by fitting to be $91\pm{1}\%$. This reduced coherence is due to imperfections in our experiment, including imperfect calibration and decoherence during the protocol. Nevertheless we observe clear evidence for high coherence of the superposition.

The increasing sensitivity and accessibility of the squeezed-Fock states as a function of both squeezing and the excitation number means that these states offer possibilities for achieving sensitivities exceeding that of either the energy eigenbasis or the squeezed vacuum alone. This could offer advantages e.g.\ for tests of quantum gravity theories \cite{17Bosso}. Combining the preparation of squeezed Fock superposition states with optical pumping of the spin would also allow the production of mixed states of the squeezed basis, with possible applications in demonstrations of the enhancement of thermodynamic processes \cite{14Rossnagel}.

The methods presented in this Letter can be easily adapted to prepare displaced-squeezed Fock states by adding a carrier tone to the used Hamiltonians \cite{15Kienzler}. The ability to choose the basis in which Jaynes-Cummings physics can be applied adds significant flexibility to the manipulation of quantum oscillators, although further theoretical work is needed to identify the optimal use of these methods. Non-Gaussian and squeezed states have been proposed previously as a resource for continuous-variable quantum computation \cite{01Gottesman, 16Terhal} and in hybrid continuous-discrete approaches
\cite{15Andersen}.  Our methods provide new possibilities for creating, controlling and measuring states with similar features.

\begin{acknowledgments}
We thank Ludwig de Clercq and Ben Keitch for contributions to the experimental apparatus. We acknowledge support from the Swiss National Science Foundation under grant no. 200021\_134776, ETH Research Grant under grant no. ETH-18 12-2, and from the Swiss National Science Foundation through the National Centre of Competence in Research for Quantum Science and Technology (QSIT). D.K.\ acknowledges support from the Swiss National Science Foundation under grant no. 165208. We thank the developers of QuTiP \cite{Johansson2013}, which was used for numerical simulations.
\end{acknowledgments}

\bibliography{myrefs}

\begin{thebibliography}{22}%
\makeatletter
\providecommand \@ifxundefined [1]{%
 \@ifx{#1\undefined}
}%
\providecommand \@ifnum [1]{%
 \ifnum #1\expandafter \@firstoftwo
 \else \expandafter \@secondoftwo
 \fi
}%
\providecommand \@ifx [1]{%
 \ifx #1\expandafter \@firstoftwo
 \else \expandafter \@secondoftwo
 \fi
}%
\providecommand \natexlab [1]{#1}%
\providecommand \enquote  [1]{``#1''}%
\providecommand \bibnamefont  [1]{#1}%
\providecommand \bibfnamefont [1]{#1}%
\providecommand \citenamefont [1]{#1}%
\providecommand \href@noop [0]{\@secondoftwo}%
\providecommand \href [0]{\begingroup \@sanitize@url \@href}%
\providecommand \@href[1]{\@@startlink{#1}\@@href}%
\providecommand \@@href[1]{\endgroup#1\@@endlink}%
\providecommand \@sanitize@url [0]{\catcode `\\12\catcode `\$12\catcode
  `\&12\catcode `\#12\catcode `\^12\catcode `\_12\catcode `\%12\relax}%
\providecommand \@@startlink[1]{}%
\providecommand \@@endlink[0]{}%
\providecommand \url  [0]{\begingroup\@sanitize@url \@url }%
\providecommand \@url [1]{\endgroup\@href {#1}{\urlprefix }}%
\providecommand \urlprefix  [0]{URL }%
\providecommand \Eprint [0]{\href }%
\providecommand \doibase [0]{http://dx.doi.org/}%
\providecommand \selectlanguage [0]{\@gobble}%
\providecommand \bibinfo  [0]{\@secondoftwo}%
\providecommand \bibfield  [0]{\@secondoftwo}%
\providecommand \translation [1]{[#1]}%
\providecommand \BibitemOpen [0]{}%
\providecommand \bibitemStop [0]{}%
\providecommand \bibitemNoStop [0]{.\EOS\space}%
\providecommand \EOS [0]{\spacefactor3000\relax}%
\providecommand \BibitemShut  [1]{\csname bibitem#1\endcsname}%
\let\auto@bib@innerbib\@empty
\bibitem [{\citenamefont {Haroche}(2013)}]{13Haroche}%
  \BibitemOpen
  \bibfield  {author} {\bibinfo {author} {\bibfnamefont {S.}~\bibnamefont
  {Haroche}},\ }\href {\doibase 10.1103/RevModPhys.85.1083} {\bibfield
  {journal} {\bibinfo  {journal} {Rev. Mod. Phys.}\ }\textbf {\bibinfo {volume}
  {85}},\ \bibinfo {pages} {1083} (\bibinfo {year} {2013})}\BibitemShut
  {NoStop}%
\bibitem [{\citenamefont {Wineland}(2013)}]{13Wineland}%
  \BibitemOpen
  \bibfield  {author} {\bibinfo {author} {\bibfnamefont {D.~J.}\ \bibnamefont
  {Wineland}},\ }\href {\doibase 10.1103/RevModPhys.85.1103} {\bibfield
  {journal} {\bibinfo  {journal} {Rev. Mod. Phys.}\ }\textbf {\bibinfo {volume}
  {85}},\ \bibinfo {pages} {1103} (\bibinfo {year} {2013})}\BibitemShut
  {NoStop}%
\bibitem [{\citenamefont {Haroche}\ and\ \citenamefont
  {Raimond}(2006)}]{BkHaroche}%
  \BibitemOpen
  \bibfield  {author} {\bibinfo {author} {\bibfnamefont {S.}~\bibnamefont
  {Haroche}}\ and\ \bibinfo {author} {\bibfnamefont {J.-M.}\ \bibnamefont
  {Raimond}},\ }\href@noop {} {\emph {\bibinfo {title} {Exploring the Quantum:
  Atoms and Cavities and Photons}}}\ (\bibinfo  {publisher} {Oxford University
  Press},\ \bibinfo {year} {(2006)})\BibitemShut {NoStop}%
\bibitem [{\citenamefont {Meekhof}\ \emph {et~al.}(1996)\citenamefont
  {Meekhof}, \citenamefont {Monroe}, \citenamefont {King}, \citenamefont
  {Itano},\ and\ \citenamefont {Wineland}}]{96Meekhof}%
  \BibitemOpen
  \bibfield  {author} {\bibinfo {author} {\bibfnamefont {D.~M.}\ \bibnamefont
  {Meekhof}}, \bibinfo {author} {\bibfnamefont {C.}~\bibnamefont {Monroe}},
  \bibinfo {author} {\bibfnamefont {B.~E.}\ \bibnamefont {King}}, \bibinfo
  {author} {\bibfnamefont {W.~M.}\ \bibnamefont {Itano}}, \ and\ \bibinfo
  {author} {\bibfnamefont {D.~J.}\ \bibnamefont {Wineland}},\ }\href@noop {}
  {\bibfield  {journal} {\bibinfo  {journal} {Phys. Rev. Lett.}\ }\textbf
  {\bibinfo {volume} {76}},\ \bibinfo {pages} {1796} (\bibinfo {year}
  {1996})},\ \bibinfo {note} {phys. Rev. Lett. 77, 2346(E) 1996}\BibitemShut
  {NoStop}%
\bibitem [{\citenamefont {Schmidt-Kaler}\ \emph {et~al.}(2003)\citenamefont
  {Schmidt-Kaler}, \citenamefont {H{\"a}ffner}, \citenamefont {Riebe},
  \citenamefont {Gulde}, \citenamefont {Lancaster}, \citenamefont {Deuschle},
  \citenamefont {Becher}, \citenamefont {Eschner},\ and\ \citenamefont
  {Blatt}}]{03SchmidtKaler}%
  \BibitemOpen
  \bibfield  {author} {\bibinfo {author} {\bibfnamefont {F.}~\bibnamefont
  {Schmidt-Kaler}}, \bibinfo {author} {\bibfnamefont {H.}~\bibnamefont
  {H{\"a}ffner}}, \bibinfo {author} {\bibfnamefont {M.}~\bibnamefont {Riebe}},
  \bibinfo {author} {\bibfnamefont {S.}~\bibnamefont {Gulde}}, \bibinfo
  {author} {\bibfnamefont {G.~P.~T.}\ \bibnamefont {Lancaster}}, \bibinfo
  {author} {\bibfnamefont {T.}~\bibnamefont {Deuschle}}, \bibinfo {author}
  {\bibfnamefont {C.}~\bibnamefont {Becher}}, \bibinfo {author} {\bibfnamefont
  {C.~F. R.~J.}\ \bibnamefont {Eschner}}, \ and\ \bibinfo {author}
  {\bibfnamefont {R.}~\bibnamefont {Blatt}},\ }\href@noop {} {\bibfield
  {journal} {\bibinfo  {journal} {Nature}\ }\textbf {\bibinfo {volume} {422}},\
  \bibinfo {pages} {408 } (\bibinfo {year} {2003})}\BibitemShut {NoStop}%
\bibitem [{\citenamefont {Law}\ and\ \citenamefont {Eberly}(1996)}]{96Law}%
  \BibitemOpen
  \bibfield  {author} {\bibinfo {author} {\bibfnamefont {C.~K.}\ \bibnamefont
  {Law}}\ and\ \bibinfo {author} {\bibfnamefont {J.~H.}\ \bibnamefont
  {Eberly}},\ }\href {\doibase 10.1103/PhysRevLett.76.1055} {\bibfield
  {journal} {\bibinfo  {journal} {Phys. Rev. Lett.}\ }\textbf {\bibinfo
  {volume} {76}},\ \bibinfo {pages} {1055} (\bibinfo {year}
  {1996})}\BibitemShut {NoStop}%
\bibitem [{\citenamefont {Ben-Kish}\ \emph {et~al.}(2003)\citenamefont
  {Ben-Kish}, \citenamefont {DeMarco}, \citenamefont {Meyer}, \citenamefont
  {Rowe}, \citenamefont {Britton}, \citenamefont {Itano}, \citenamefont
  {Jelenkovi\ifmmode~\acute{c}\else \'{c}\fi{}}, \citenamefont {Langer},
  \citenamefont {Leibfried}, \citenamefont {Rosenband},\ and\ \citenamefont
  {Wineland}}]{03BenKish}%
  \BibitemOpen
  \bibfield  {author} {\bibinfo {author} {\bibfnamefont {A.}~\bibnamefont
  {Ben-Kish}}, \bibinfo {author} {\bibfnamefont {B.}~\bibnamefont {DeMarco}},
  \bibinfo {author} {\bibfnamefont {V.}~\bibnamefont {Meyer}}, \bibinfo
  {author} {\bibfnamefont {M.}~\bibnamefont {Rowe}}, \bibinfo {author}
  {\bibfnamefont {J.}~\bibnamefont {Britton}}, \bibinfo {author} {\bibfnamefont
  {W.~M.}\ \bibnamefont {Itano}}, \bibinfo {author} {\bibfnamefont {B.~M.}\
  \bibnamefont {Jelenkovi\ifmmode~\acute{c}\else \'{c}\fi{}}}, \bibinfo
  {author} {\bibfnamefont {C.}~\bibnamefont {Langer}}, \bibinfo {author}
  {\bibfnamefont {D.}~\bibnamefont {Leibfried}}, \bibinfo {author}
  {\bibfnamefont {T.}~\bibnamefont {Rosenband}}, \ and\ \bibinfo {author}
  {\bibfnamefont {D.~J.}\ \bibnamefont {Wineland}},\ }\href {\doibase
  10.1103/PhysRevLett.90.037902} {\bibfield  {journal} {\bibinfo  {journal}
  {Phys. Rev. Lett.}\ }\textbf {\bibinfo {volume} {90}},\ \bibinfo {pages}
  {037902} (\bibinfo {year} {2003})}\BibitemShut {NoStop}%
\bibitem [{\citenamefont {Hofheinz}\ \emph {et~al.}(2009)\citenamefont
  {Hofheinz}, \citenamefont {Wang}, \citenamefont {Ansmann}, \citenamefont
  {Bialczak}, \citenamefont {Lucero}, \citenamefont {Neeley}, \citenamefont
  {O'Connell}, \citenamefont {Sank}, \citenamefont {Wenner}, \citenamefont
  {Martinis},\ and\ \citenamefont {Cleland}}]{09Hofheinz}%
  \BibitemOpen
  \bibfield  {author} {\bibinfo {author} {\bibfnamefont {M.}~\bibnamefont
  {Hofheinz}}, \bibinfo {author} {\bibfnamefont {H.}~\bibnamefont {Wang}},
  \bibinfo {author} {\bibfnamefont {M.}~\bibnamefont {Ansmann}}, \bibinfo
  {author} {\bibfnamefont {R.~C.}\ \bibnamefont {Bialczak}}, \bibinfo {author}
  {\bibfnamefont {E.}~\bibnamefont {Lucero}}, \bibinfo {author} {\bibfnamefont
  {M.}~\bibnamefont {Neeley}}, \bibinfo {author} {\bibfnamefont {A.~D.}\
  \bibnamefont {O'Connell}}, \bibinfo {author} {\bibfnamefont {D.}~\bibnamefont
  {Sank}}, \bibinfo {author} {\bibfnamefont {J.}~\bibnamefont {Wenner}},
  \bibinfo {author} {\bibfnamefont {J.~M.}\ \bibnamefont {Martinis}}, \ and\
  \bibinfo {author} {\bibfnamefont {A.~N.}\ \bibnamefont {Cleland}},\ }\href
  {\doibase doi:10.1038/nature08005} {\bibfield  {journal} {\bibinfo  {journal}
  {Nature}\ }\textbf {\bibinfo {volume} {459}},\ \bibinfo {pages} {546}
  (\bibinfo {year} {2009})}\BibitemShut {NoStop}%
\bibitem [{\citenamefont {Kienzler}\ \emph {et~al.}(2015)\citenamefont
  {Kienzler}, \citenamefont {Lo}, \citenamefont {Keitch}, \citenamefont
  {de~Clercq}, \citenamefont {Leupold}, \citenamefont {Lindenfelser},
  \citenamefont {Marinelli}, \citenamefont {Negnevitsky},\ and\ \citenamefont
  {Home}}]{15Kienzler}%
  \BibitemOpen
  \bibfield  {author} {\bibinfo {author} {\bibfnamefont {D.}~\bibnamefont
  {Kienzler}}, \bibinfo {author} {\bibfnamefont {H.-Y.}\ \bibnamefont {Lo}},
  \bibinfo {author} {\bibfnamefont {B.}~\bibnamefont {Keitch}}, \bibinfo
  {author} {\bibfnamefont {L.}~\bibnamefont {de~Clercq}}, \bibinfo {author}
  {\bibfnamefont {F.}~\bibnamefont {Leupold}}, \bibinfo {author} {\bibfnamefont
  {F.}~\bibnamefont {Lindenfelser}}, \bibinfo {author} {\bibfnamefont
  {M.}~\bibnamefont {Marinelli}}, \bibinfo {author} {\bibfnamefont
  {V.}~\bibnamefont {Negnevitsky}}, \ and\ \bibinfo {author} {\bibfnamefont
  {J.~P.}\ \bibnamefont {Home}},\ }\href {\doibase 10.1126/science.1261033}
  {\bibfield  {journal} {\bibinfo  {journal} {Science}\ }\textbf {\bibinfo
  {volume} {347}},\ \bibinfo {pages} {53} (\bibinfo {year} {2015})}\BibitemShut
  {NoStop}%
\bibitem [{\citenamefont {Kienzler}\ \emph {et~al.}(2016)\citenamefont
  {Kienzler}, \citenamefont {Fl\"uhmann}, \citenamefont {Negnevitsky},
  \citenamefont {Lo}, \citenamefont {Marinelli}, \citenamefont {Nadlinger},\
  and\ \citenamefont {Home}}]{16Kienzler}%
  \BibitemOpen
  \bibfield  {author} {\bibinfo {author} {\bibfnamefont {D.}~\bibnamefont
  {Kienzler}}, \bibinfo {author} {\bibfnamefont {C.}~\bibnamefont
  {Fl\"uhmann}}, \bibinfo {author} {\bibfnamefont {V.}~\bibnamefont
  {Negnevitsky}}, \bibinfo {author} {\bibfnamefont {H.-Y.}\ \bibnamefont {Lo}},
  \bibinfo {author} {\bibfnamefont {M.}~\bibnamefont {Marinelli}}, \bibinfo
  {author} {\bibfnamefont {D.}~\bibnamefont {Nadlinger}}, \ and\ \bibinfo
  {author} {\bibfnamefont {J.~P.}\ \bibnamefont {Home}},\ }\href {\doibase
  10.1103/PhysRevLett.116.140402} {\bibfield  {journal} {\bibinfo  {journal}
  {Phys. Rev. Lett.}\ }\textbf {\bibinfo {volume} {116}},\ \bibinfo {pages}
  {140402} (\bibinfo {year} {2016})}\BibitemShut {NoStop}%
\bibitem [{\citenamefont {Plebanski}(1956)}]{56Plebanski}%
  \BibitemOpen
  \bibfield  {author} {\bibinfo {author} {\bibfnamefont {J.}~\bibnamefont
  {Plebanski}},\ }\href {\doibase 10.1103/PhysRev.101.1825} {\bibfield
  {journal} {\bibinfo  {journal} {Phys. Rev.}\ }\textbf {\bibinfo {volume}
  {101}},\ \bibinfo {pages} {1825} (\bibinfo {year} {1956})}\BibitemShut
  {NoStop}%
\bibitem [{\citenamefont {Satyanarayana}(1985)}]{85Satyanarayana}%
  \BibitemOpen
  \bibfield  {author} {\bibinfo {author} {\bibfnamefont {M.~V.}\ \bibnamefont
  {Satyanarayana}},\ }\href {\doibase 10.1103/PhysRevD.32.400} {\bibfield
  {journal} {\bibinfo  {journal} {Phys. Rev. D}\ }\textbf {\bibinfo {volume}
  {32}},\ \bibinfo {pages} {400} (\bibinfo {year} {1985})}\BibitemShut
  {NoStop}%
\bibitem [{\citenamefont {Kral}(1990)}]{90Kral}%
  \BibitemOpen
  \bibfield  {author} {\bibinfo {author} {\bibfnamefont {P.}~\bibnamefont
  {Kral}},\ }\href@noop {} {\bibfield  {journal} {\bibinfo  {journal} {J. Mod.
  Opt.}\ }\textbf {\bibinfo {volume} {37}},\ \bibinfo {pages} {889} (\bibinfo
  {year} {1990})}\BibitemShut {NoStop}%
\bibitem [{\citenamefont {Nieto}(1997)}]{97Nieto}%
  \BibitemOpen
  \bibfield  {author} {\bibinfo {author} {\bibfnamefont {M.~M.}\ \bibnamefont
  {Nieto}},\ }\href {\doibase http://dx.doi.org/10.1016/S0375-9601(97)00183-7}
  {\bibfield  {journal} {\bibinfo  {journal} {Physics Letters A}\ }\textbf
  {\bibinfo {volume} {229}},\ \bibinfo {pages} {135 } (\bibinfo {year}
  {1997})}\BibitemShut {NoStop}%
\bibitem [{\citenamefont {Poyatos}\ \emph {et~al.}(1996)\citenamefont
  {Poyatos}, \citenamefont {Cirac},\ and\ \citenamefont {Zoller}}]{96Poyatos}%
  \BibitemOpen
  \bibfield  {author} {\bibinfo {author} {\bibfnamefont {J.~F.}\ \bibnamefont
  {Poyatos}}, \bibinfo {author} {\bibfnamefont {J.~I.}\ \bibnamefont {Cirac}},
  \ and\ \bibinfo {author} {\bibfnamefont {P.}~\bibnamefont {Zoller}},\ }\href
  {\doibase 10.1103/PhysRevLett.77.4728} {\bibfield  {journal} {\bibinfo
  {journal} {Phys. Rev. Lett.}\ }\textbf {\bibinfo {volume} {77}},\ \bibinfo
  {pages} {4728} (\bibinfo {year} {1996})}\BibitemShut {NoStop}%
\bibitem [{\citenamefont {Turchette}\ \emph {et~al.}(2000)\citenamefont
  {Turchette}, \citenamefont {Myatt}, \citenamefont {King}, \citenamefont
  {Sackett}, \citenamefont {Kielpinski}, \citenamefont {Itano}, \citenamefont
  {Monroe},\ and\ \citenamefont {Wineland}}]{00Turchette2}%
  \BibitemOpen
  \bibfield  {author} {\bibinfo {author} {\bibfnamefont {Q.~A.}\ \bibnamefont
  {Turchette}}, \bibinfo {author} {\bibfnamefont {C.~J.}\ \bibnamefont
  {Myatt}}, \bibinfo {author} {\bibfnamefont {B.~E.}\ \bibnamefont {King}},
  \bibinfo {author} {\bibfnamefont {C.~A.}\ \bibnamefont {Sackett}}, \bibinfo
  {author} {\bibfnamefont {D.}~\bibnamefont {Kielpinski}}, \bibinfo {author}
  {\bibfnamefont {W.~M.}\ \bibnamefont {Itano}}, \bibinfo {author}
  {\bibfnamefont {C.}~\bibnamefont {Monroe}}, \ and\ \bibinfo {author}
  {\bibfnamefont {D.~J.}\ \bibnamefont {Wineland}},\ }\href {\doibase
  10.1103/PhysRevA.62.053807} {\bibfield  {journal} {\bibinfo  {journal} {Phys.
  Rev. A}\ }\textbf {\bibinfo {volume} {62}},\ \bibinfo {pages} {053807}
  (\bibinfo {year} {2000})}\BibitemShut {NoStop}%
\bibitem [{\citenamefont {Bosso}\ \emph {et~al.}(2017)\citenamefont {Bosso},
  \citenamefont {Das},\ and\ \citenamefont {Mann}}]{17Bosso}%
  \BibitemOpen
  \bibfield  {author} {\bibinfo {author} {\bibfnamefont {P.}~\bibnamefont
  {Bosso}}, \bibinfo {author} {\bibfnamefont {S.}~\bibnamefont {Das}}, \ and\
  \bibinfo {author} {\bibfnamefont {R.~B.}\ \bibnamefont {Mann}},\ }\href@noop
  {} {\  (\bibinfo {year} {2017})},\ \Eprint {http://arxiv.org/abs/1704.08198}
  {arXiv:1704.08198 [gr-qc]} \BibitemShut {NoStop}%
\bibitem [{\citenamefont {Ro\ss{}nagel}\ \emph {et~al.}(2014)\citenamefont
  {Ro\ss{}nagel}, \citenamefont {Abah}, \citenamefont {Schmidt-Kaler},
  \citenamefont {Singer},\ and\ \citenamefont {Lutz}}]{14Rossnagel}%
  \BibitemOpen
  \bibfield  {author} {\bibinfo {author} {\bibfnamefont {J.}~\bibnamefont
  {Ro\ss{}nagel}}, \bibinfo {author} {\bibfnamefont {O.}~\bibnamefont {Abah}},
  \bibinfo {author} {\bibfnamefont {F.}~\bibnamefont {Schmidt-Kaler}}, \bibinfo
  {author} {\bibfnamefont {K.}~\bibnamefont {Singer}}, \ and\ \bibinfo {author}
  {\bibfnamefont {E.}~\bibnamefont {Lutz}},\ }\href {\doibase
  10.1103/PhysRevLett.112.030602} {\bibfield  {journal} {\bibinfo  {journal}
  {Phys. Rev. Lett.}\ }\textbf {\bibinfo {volume} {112}},\ \bibinfo {pages}
  {030602} (\bibinfo {year} {2014})}\BibitemShut {NoStop}%
\bibitem [{\citenamefont {Gottesman}\ \emph {et~al.}(2001)\citenamefont
  {Gottesman}, \citenamefont {Kitaev},\ and\ \citenamefont
  {Preskill}}]{01Gottesman}%
  \BibitemOpen
  \bibfield  {author} {\bibinfo {author} {\bibfnamefont {D.}~\bibnamefont
  {Gottesman}}, \bibinfo {author} {\bibfnamefont {A.}~\bibnamefont {Kitaev}}, \
  and\ \bibinfo {author} {\bibfnamefont {J.}~\bibnamefont {Preskill}},\ }\href
  {\doibase 10.1103/PhysRevA.64.012310} {\bibfield  {journal} {\bibinfo
  {journal} {Phys. Rev. A}\ }\textbf {\bibinfo {volume} {64}},\ \bibinfo
  {pages} {012310} (\bibinfo {year} {2001})}\BibitemShut {NoStop}%
\bibitem [{\citenamefont {Terhal}\ and\ \citenamefont
  {Weigand}(2016)}]{16Terhal}%
  \BibitemOpen
  \bibfield  {author} {\bibinfo {author} {\bibfnamefont {B.~M.}\ \bibnamefont
  {Terhal}}\ and\ \bibinfo {author} {\bibfnamefont {D.}~\bibnamefont
  {Weigand}},\ }\href {\doibase 10.1103/PhysRevA.93.012315} {\bibfield
  {journal} {\bibinfo  {journal} {Phys. Rev. A}\ }\textbf {\bibinfo {volume}
  {93}},\ \bibinfo {pages} {012315} (\bibinfo {year} {2016})}\BibitemShut
  {NoStop}%
\bibitem [{\citenamefont {Andersen}\ \emph {et~al.}(2015)\citenamefont
  {Andersen}, \citenamefont {Neergaard-Nielsen}, \citenamefont {van Loock},\
  and\ \citenamefont {Furusawa}}]{15Andersen}%
  \BibitemOpen
  \bibfield  {author} {\bibinfo {author} {\bibfnamefont {U.~L.}\ \bibnamefont
  {Andersen}}, \bibinfo {author} {\bibfnamefont {J.~S.}\ \bibnamefont
  {Neergaard-Nielsen}}, \bibinfo {author} {\bibfnamefont {P.}~\bibnamefont {van
  Loock}}, \ and\ \bibinfo {author} {\bibfnamefont {A.}~\bibnamefont
  {Furusawa}},\ }\href {http://dx.doi.org/10.1038/nphys3410} {\bibfield
  {journal} {\bibinfo  {journal} {Nat Phys}\ }\textbf {\bibinfo {volume}
  {11}},\ \bibinfo {pages} {713} (\bibinfo {year} {2015})}\BibitemShut
  {NoStop}%
\bibitem [{\citenamefont {Johansson}\ \emph {et~al.}(2013)\citenamefont
  {Johansson}, \citenamefont {Nation},\ and\ \citenamefont
  {Nori}}]{Johansson2013}%
  \BibitemOpen
  \bibfield  {author} {\bibinfo {author} {\bibfnamefont {J.}~\bibnamefont
  {Johansson}}, \bibinfo {author} {\bibfnamefont {P.}~\bibnamefont {Nation}}, \
  and\ \bibinfo {author} {\bibfnamefont {F.}~\bibnamefont {Nori}},\ }\href
  {\doibase 10.1016/j.cpc.2012.11.019} {\bibfield  {journal} {\bibinfo
  {journal} {Computer Physics Communications}\ }\textbf {\bibinfo {volume}
  {184}},\ \bibinfo {pages} {1234} (\bibinfo {year} {2013})}\BibitemShut
  {NoStop}%
\end{thebibliography}%

\appendix

\section*{Supplemental Material}
\begin{figure}[t]
	\centering
	\resizebox{0.49 \textwidth}{!}{\includegraphics{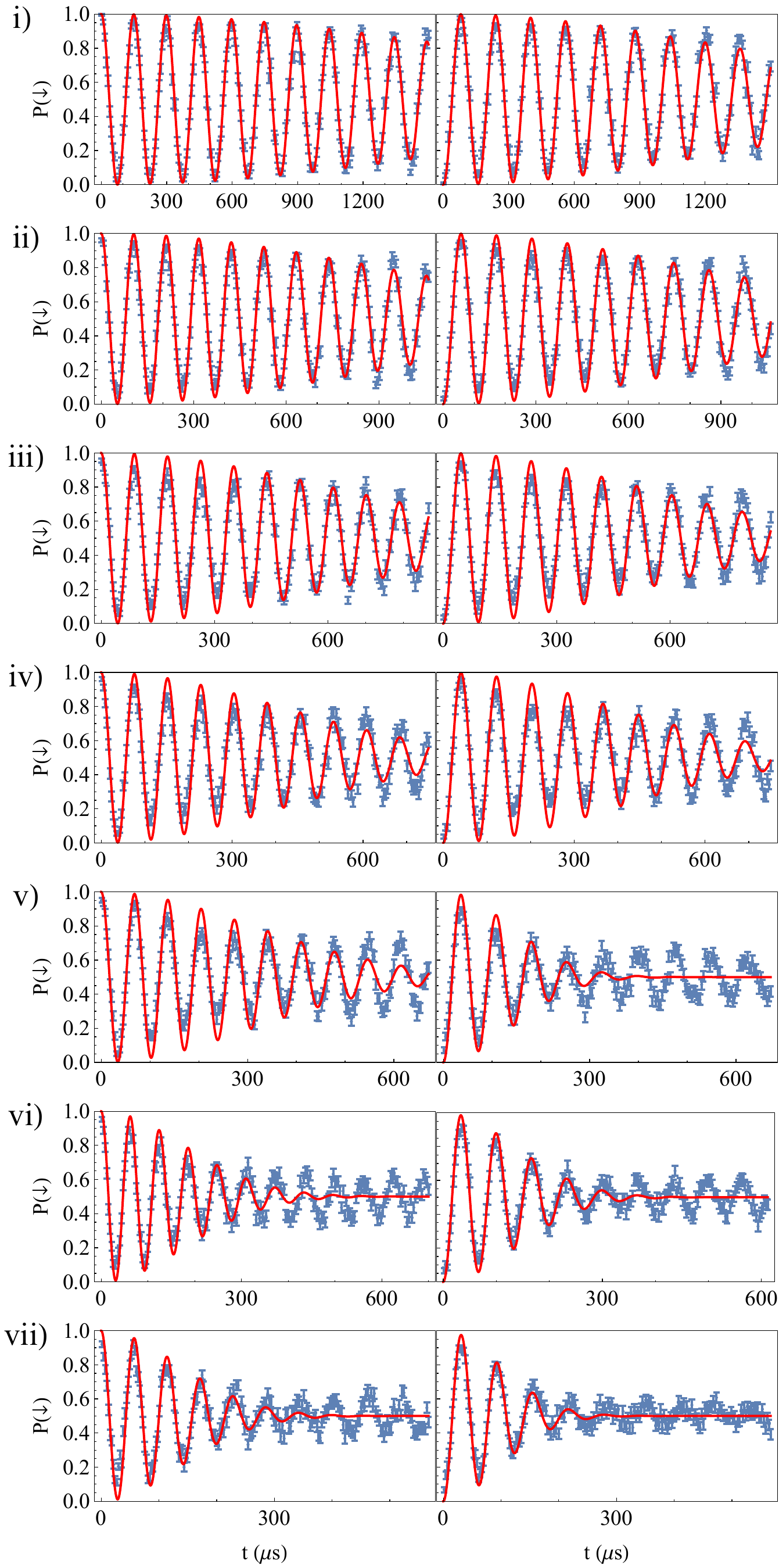}}
	\caption{Rabi oscillation data for the $\hat{H}_+$ and $\hat{H}_-$ transitions for each of the transitions a) $\ket{\downarrow (\uparrow)}\ket{\zeta, n} \leftrightarrow \ket{\uparrow (\downarrow)} \ket{\zeta, n + 1}$ for $n$ from 0 to 6, given as plots i) to vii) respectively. Data shown on the left is for $\hat{H}_+$ and on the right are the data for $\hat{H}_-$. The fits are discussed in the text of this section. }
	\label{fig:timescans}
\end{figure}

In figure \ref{fig:timescans} we give the spin population data for the $\hat{H}_{+}$ and $\hat{H}_-$ Rabi oscillations. In this supplement we discuss the effects which we think are playing the primary role in the loss of oscillation contrast in these experiments, which we ascribe to time-evolution of the basis states of the system due to a combination of trap frequency offsets and fast fluctuations. This is different to standard Jaynes-Cummings physics for which the basis states (the energy eigenstates) are themselves immune to dephasing (although dephasing of the superposition of the two energy eigenstates during the Rabi oscillations does play a role). The squeezed Fock states are sensitive to this mechanism because their axis of squeezing evolves with a rate given by this detuning. This sensitivity also increases with excitation number $n$.

The data which were used to produce the scaling of the Rabi frequencies in figure 2 of the main text are shown in figure \ref{fig:timescans}. For higher $n$, it is clear that the loss of coherence does not fit well to either a Gaussian or exponential - the examples in figure \ref{fig:timescans} are for Gaussian decay, fitting a function
\be
P(\downarrow, t) &=& \frac{(1 + (-1)^p)}{2} \nonumber \\
&+& (-1)^p\frac{1}{2}\left(1 + e^{-\gamma t^2} \cos(\Omega t) \right)
\ee
where $p = 1$ for $\hat{H}_+$ flopping and $p = 0$ for $\hat{H}_-$ flopping. Neither the Gaussian or exponential decay models are able to account for the amplitude of oscillations at both long and short times. Rabi oscillations persist at higher amplitudes at long times than would be expected from fits to the data at short times. This effect is particularly apparent in the higher $n$ states.

\begin{figure}[t]
	\centering
	\resizebox{0.49 \textwidth}{!}{\includegraphics{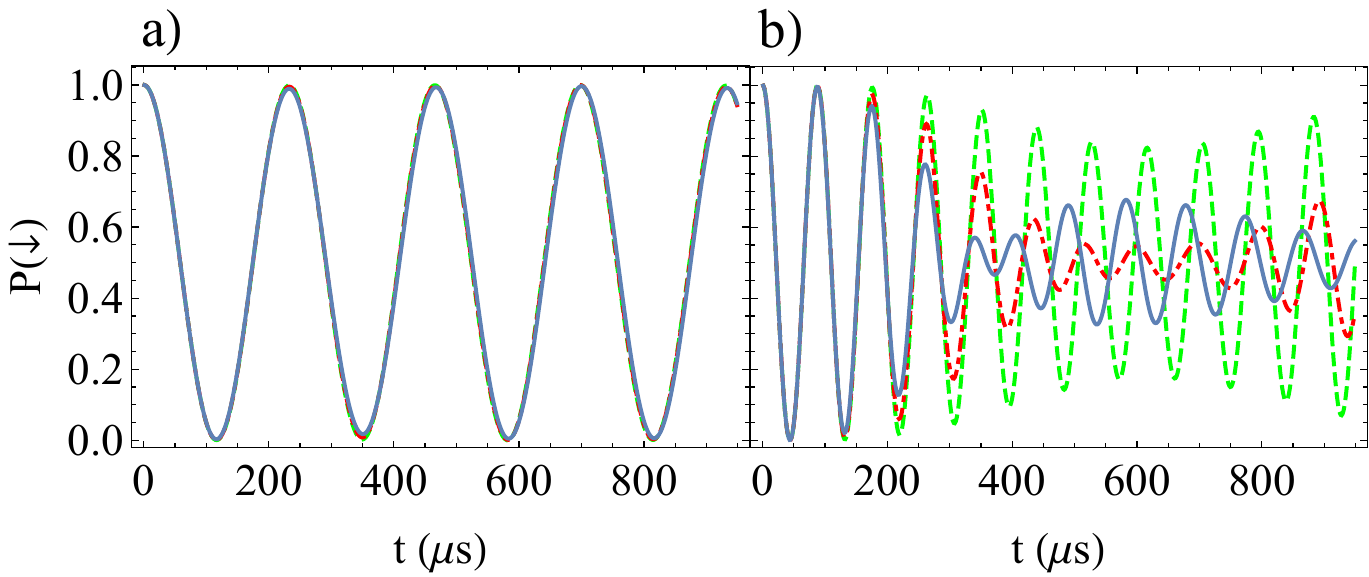}}
	\caption{Simulated flopping curves on the $\hat{H}_+$ Hamiltonian for a) $\ket{\downarrow} \ket{\zeta, 0}\leftrightarrow \ket{\uparrow} \ket{\zeta, 1}$ and b) $\ket{\downarrow} \ket{\zeta, 6}\leftrightarrow \ket{\uparrow}$ for fixed detunings of \SI{10}{\hertz} (green, dashed), \SI{20}{\hertz} (red, dot-dashed) and \SI{30}{\hertz} (blue, solid) using $r = 1$ and $\Omega/(2 \pi) =$~ \SI{4.3}{\kilo\hertz}. While for the ground state simulation no effect of the detuning is visible and the traces are nearly overlapped, for the excited state the effects of the detuning are clearly visible. In addition to a change in the Rabi oscillation frequency a collapse and revival effect is observed. This is due to the transition terms in the Hamiltonian of equation \ref{eq:hamdet}. }
	\label{fig:sim5}
\end{figure}

To understand this better, we performed simulations of the population evolution under the influence of an $\hat{H}_+$ Hamiltonian with a fixed detuning between the trap frequency and the frequency corresponding to half the difference between the red-sideband and blue-sideband drives. This would account for a mis-setting of the drive frequencies, or equivalently that the trap frequency of the ion had shifted. Figure \ref{fig:sim5} shows flopping as a function of time for  $\ket{\downarrow} \ket{\zeta, 0}\leftrightarrow \ket{\uparrow} \ket{\zeta, 1}$ and  $\ket{\downarrow} \ket{\zeta, 6}\leftrightarrow \ket{\uparrow} \ket{\zeta, 7}$ for example curves for a 10~Hz, 20~Hz and 30~Hz detuning. In simulating such systems the energy eigenstate basis is not an efficient choice. Instead, we have found it useful to consider the effect of the detuning in terms of the creation and annihilation operators in the squeezed Fock basis. The Hamiltonian $\hat{H}_{\rm det} = \hbar \delta \create \destroy$ can then be rewritten as
\be
\hat{H}_{\rm det} &=& \hbar \delta \left( \hat{K}^\dagger\hat{K} \cosh(2 r) + \sinh^2(r)\right) \nonumber \\
&+& \hbar \delta \frac{\sinh(2r)}{2} \left((\hat{K}^\dagger)^2 + \hat{K}^2\right). \label{eq:hamdet}
\ee
The first term would imply that the Jaynes-Cummings Hamiltonian does not meet the resonance condition. By analogy with standard Rabi physics (and as we see in simulations) this leads to Rabi oscillations with a higher frequency but with amplitude $<1$. The final two terms induce transitions to nearby states differing in their quantum number by $2$, with some probability during the evolution to come back to the original state. These lead to the collapse and revival effects which can be observed in the simulations shown in  figure \ref{fig:sim5}. It is notable that the $n$ dependent terms in the Hamiltonian scale as $n$ for the diagonal terms, and as $\sqrt{(n + 1)(n + 2)}$ and $\sqrt{n(n-1)}$ for the off-diagonal terms, which produces the amplified sensitivity for the higher $n$ states. In the limit of large $r$ both terms scale as $\exp(2r)$, which is proportional to the quadrature squeezing factor.

\begin{figure}
	\centering
	\resizebox{0.49 \textwidth}{!}{\includegraphics{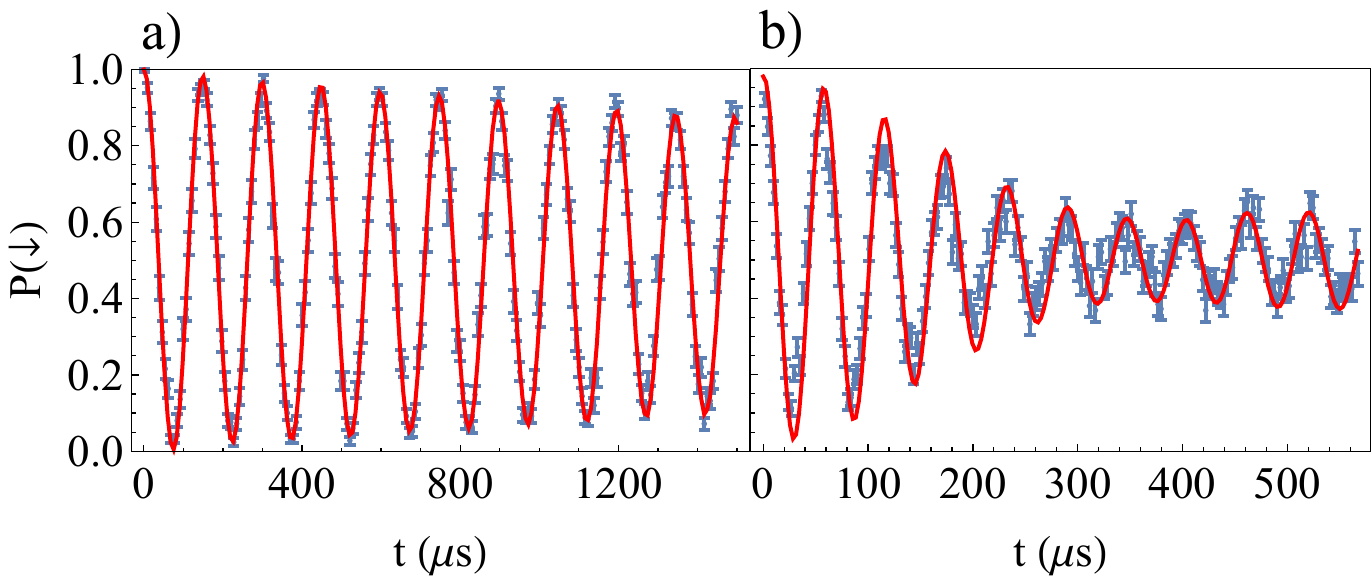}}
	\caption{Experimental data overlaid with simulated flopping curves using a Master equation approach on the $\hat{H}_+$ Hamiltonian for a) $\ket{\downarrow} \ket{\zeta, 0}\leftrightarrow \ket{\uparrow} \ket{\zeta, 1}$ and b) $\ket{\downarrow} \ket{\zeta, 6}\leftrightarrow \ket{\uparrow}$ with a fixed detuning of \SI{30}{\hertz} combined with Lindblad collapse operators for heating and dephasing noise.}
	\label{fig;simDaniel}
\end{figure}

It is noteworthy that a fixed offset might be hard to measure when close to the bottom of the state ladder, but would only become apparent at higher excitations. For each of the fixed frequency offsets we see partial collapse and revival occurring in both cases. This effect is not dissimilar to what we see in the data, although it is a weak effect. The suspicion of a fixed frequency offset is supported by the presence of data sets in which the  $\hat{H}_+$ data shows different behaviour to those of $\hat{H}_-$ even though the transitions used are between the same pair of Fock states. Since the two measurements were performed some time apart, this is perhaps due to a drift in the trap frequency between the measurements.  We expect that in addition to a possible fixed offset fast fluctuations of our trap frequency add to the decay. As an example we perform a simulation solving the Master equation in Lindblad form, in which we use the detuned versions of $\hat{H}_+$ and $\hat{H}_-$ with a fixed detuning of \SI{30}{\hertz} and jump operators for an amplitude reservoir ($\hat{L}_{\rm{A1}}=\sqrt\Gamma_{\rm{A}} \create$, $\hat{L}_{\rm{A2}}=\sqrt\Gamma_{\rm{A}} \destroy$ with $\Gamma_{\rm{A}}/(2\pi) =$ ~\SI{10.7}{Hz} calibrated by measuring the rate of heating of the motional ground state) and a phase reservoir ($\hat{L}_{\rm{Ph}}=\sqrt\Gamma_{\rm{Ph}} \create\destroy$ with $\Gamma_{\rm{Ph}}/(2\pi) =$~\SI{5}{Hz}, tuned to match the flopping curves in the squeezed state ladder) \cite{00Turchette2}. We calculate the evolution in the same stepwise fashion as in the experiment with $\hat{H}_+$ and $\hat{H}_-$ applied alternating and a final probe pulse of the respective Hamiltonian. Results of the simulation are shown in figure \ref{fig;simDaniel}, matching the data better than using any effect alone.

\newpage 
\phantom{whatever}

\end{document}